\documentclass[useAMS,usenatbib,letter]{mn2e}
\usepackage{times}
\usepackage{graphicx}
\usepackage{amssymb,amsmath}	
\usepackage{mathrsfs}
\usepackage[utf8x]{inputenc}
\usepackage[usenames]{color}
\usepackage{color}

\PrerenderUnicode{áéíóúñ}
\title[Photosphere of variable GRB jets]{Photospheric emission from
  long duration gamma-ray bursts powered by variable engines}

\author[L\'opez-C\'amara et al.] {Diego
  L\'opez-C\'amara$^1$, Brian J. Morsony$^{2}$, Davide Lazzati$^{1,3}$\thanks{E-mail:lazzatid@science.oregonstate.edu} \\
  $^{1}$Department of Physics, NC State University, 2401 Stinson
  Drive, Raleigh, NC 27695-8202, USA \\
  $^{2}$Department of Astronomy, University of Wisconsin-Madison,
  2535 Sterling Hall, 475 N. Charter Street, Madison WI 53706-1582, USA \\
  $^{3}$Department of Physics, Oregon State University, 301 Weniger
  Hall, Corvallis, OR 97331, USA}

\begin{document}

\maketitle
\label{firstpage}

\begin{abstract}
  We present the results of a set of numerical simulations of
  long-duration gamma-ray burst jets aimed at studying the effect of a
  variable engine on the peak frequency of the photospheric
  emission. Our simulations follow the propagation of the jet inside
  the progenitor star, its break-out, and the subsequent expansion in
  the environment out to the photospheric radius. A constant and two
  step-function models are considered for the engine luminosity. We
  show that our synthetic light-curves follow a luminosity-peak
  frequency correlation analogous to the Golenetskii correlation found
  in long-duration gamma-ray burst observations. Within the parameter
  space explored, it appears that the central engine luminosity
  profile does not have a significant effect on the location of a
  gamma-ray burst in the Luminosity-peak frequency plane, bursts from
  different central engines being indistinguishable from each other.
\end{abstract}

\begin{keywords}
gamma-ray bursts: general --- radiation mechanisms: thermal --- hydrodynamics
\end{keywords} 

\section{Introduction}\label{sec:intro}
Ever since the detection of the first Gamma-ray burst (GRB) by
\citet{klebesadel73} and with the increase of the number of observed
GRBs it has been clear that many of them share some general
characteristics and so have even been grouped together in
sub-classification groups. For example, depending on their duration
GRBs have been classified in either long or short
\citep{kouveliotou93}. Strikingly, there are no two GRBs which are
exactly the same as the other. Variability is commonly observed
\citep{walker00} in GRBs, and a significant fraction of the long GRBs
($\sim$85\%) appear to be the result of several pulses
\citep{borgonovo07}. The pre- and post-bursting activity, as well as
dormant periods, still remain to be fully understood
\citep{drago07}. \citet{fenimore00} discovered a correlation between
the variability and the observed peak isotropic luminosity. Thus, it
is noteworthy to study the effects that a pulsed central engine has on
the prompt GRB emission.

The prompt emission of GRBs is characterized by bright, non-thermal
spectra, peaking between a few tens of keV up to several MeV
\citep{band93,kaneko06,gruber14}. The radiation mechanism responsible for the
production of such emission is not fully understood, possibly owing to
the great diversity of GRB spectra and light curves. Even though most
proposed models are capable of finding a parameter set to fit any GRB
spectrum, it has been so far impossible to make a synthesis and
formulate a model that can successfully account for the diversity of
the observations without requiring an adaptation to each individual
burst, and resorting to extreme fine-tuning in some cases. An important
tool in the effort of finding common properties among the diversity of
burst observations is the sample of correlations among different
bursts, such as the Amati, Yonetoku, and luminosity-Lorentz factor
correlations (see \citet{amati02, yonetoku04, ghirlanda12} for further
details, respectively).

In previous publications \citep{laz11,laz13a} we have shown that the
photospheric emission model for the prompt GRB emission can reproduce
the Amati and luminosity-Lorentz factor correlations without requiring
any fine tuning of parameters or any underlying correlation between
the properties of the central engine and/or its relativistic
outflow. Our simulations showed that the correlations are due to the
most part to the observer angle effect: burst seen close to their jet
axes appear brighter, have a higher peak frequency, and are produced
by faster ejecta compared to bursts observed near the edge of their
jets. One lingering uncertainty was, however, the robustness of the
observational correlations that we attempted to explain. A significant
amount of work has been carried out in trying to establish the role of
selection effects in the Amati and Yonetoku correlations, with
contradictory results, at best
\citep{band05,nakar05,ghirlanda08,butler09,krimm09,kocevski12,heussaff13}. A
more robust correlation that is certainly not affected by selection
effects is the Golenetskii correlation, discovered by the Konus
experiment \citep{golenetskii83} and confirmed more recently with the
high-quality Fermi spectral data \citep{ghirlanda10,lu12}.  According
to the Golenetskii correlation, different time intervals of a single
burst aligned along a straight line when plotted in the
luminosity-peak frequency plane (for further discussion see
\citet{bhat94, borgonovo01, ford95, kargatis94, lu10, norris86,
  peng09}. The burst spectrum peaks at lower frequencies when the
emission is weak but moves to higher frequencies when it is bright.
In order to study whether photosphere-dominated bursts obey this
correlation, we have carried out three hydrodynamic simulations of
relativistic jets from collapsars. Two have engines with highly
variable energy output while the third, the control, has a constant
engine, analogously to our previous work \citep{mor10}. We note that
the fact that photosphere-dominated GRBs obey the Amati correlation
does not imply that they should be able to reproduce the Golenetskii
correlation. As a matter of fact, we showed that the difference in
viewing angle is fundamental in producing the Amati correlation within
the photospheric scenario \citep{laz11,laz13a}. In the Golenetskii
case, instead, the viewing angle cannot play any role, since the
observer is the same throughout the burst. This paper is organized as
follows. We first describe the initial setup, and the numerical models
in Section~\ref{sec:input}, followed by discussion of the the
morphology, photospheric, and observable correlation's from our models
in Section~\ref{sec:results}..  Conclusions are given in
Section~\ref{sec:conc}.

\section{Initial setup and numerical models}\label{sec:input}
A variable relativistic two-dimensional (2D) jet was followed as it
drilled through the stellar progenitor, and then as it evolved through
an extremely large interstellar (ISM) domain. The jet was followed at
all times with comparable resolution ($\Delta$=8$\times$10$^{6}$~cm)
as in two-dimensional previous studies \citep{zwm03, zwh04, mor07,
  mor10, laz09, laz12, nag11, miz13}, in a domain large enough to be
able to include the radius at which the spectrum is formed ($R_{sp}
\sim$10$^{12}$~cm). The spectrum formation radius is smaller than the
photospheric radius, at which the Thomson scattering opacity equals
unity \citep{giannios12}.  In order to solve correctly the transition
between the active and quiescent epochs, the cocoon (region which
surrounds the jet and which is conformed by a mixture of shocked jet
and shocked stellar material \citep{rcr02, laz05}) was solved with a
much finer resolution ($\Delta$=3.2$\times$10$^{7}$~cm) than the
previous numerical studies.  In order for the domain to include the
spectral radius $R_{\rm{sp}}$, the simulation box was
2.56$\times$10$^{12}$~cm in length (along the jet direction) and
5.12$\times$10$^{11}$~cm across. The difference between the finest
resolution level and the size of the domain (5 orders of magnitude),
combined with the necessary integration time for the jet to reach
$R_{\rm{sp}}$ ($\sim$100~s), required the use of an adaptive mesh
code. Thus, we used the Flash code (version 2.5) \citep{fryx00}.

\begin{figure}
\centering
    \includegraphics[width=\columnwidth]{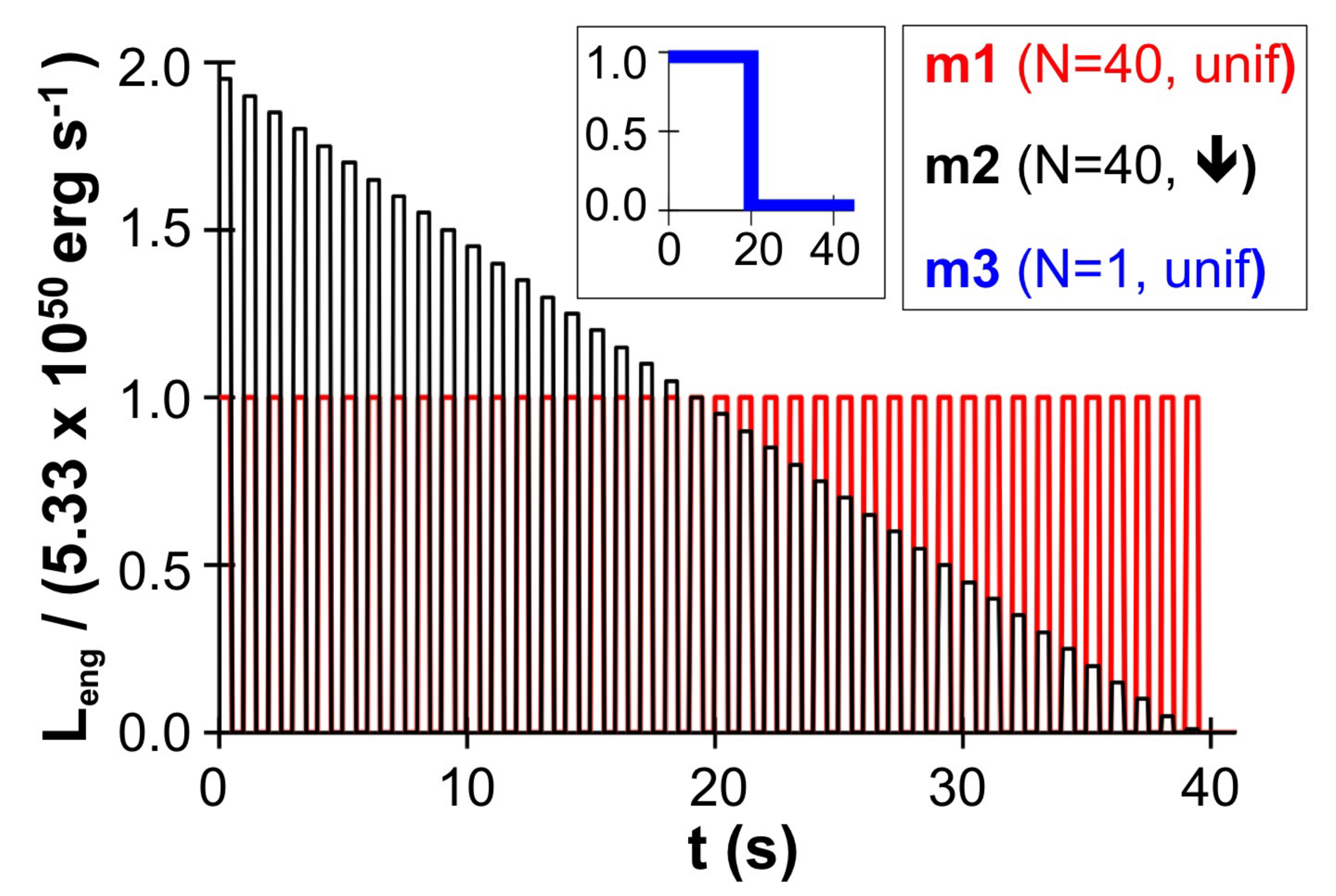}
    \caption{Engine luminosity for models m1 (red line), m2 (black line), and m3 (blue line).}
\label{fig:fig1}
\end{figure}

The numerical setup, physics, and assumptions in this study were,
unless stated differently, the same as those in \citep{laz13a}. The
stellar progenitor, model 16TI from \citep{wh06}, was immersed in a
interstellar medium with constant density
($\rho_{\rm{ism}}$=10$^{-13}$~g~cm$^{-3}$). The variable jet was
launched from the core of the progenitor and depending on the model
was followed for a total integration time between 110 and 135
seconds. This time was long enough for the jets to break out of the
progenitor, evolve through the ISM, and cross entirely the
$R_{\rm{sp}}$. The jet had at all times a half-opening angle
$\theta_0$=10$^{{o}}$ at injection, an initial Lorentz Factor
$\Gamma_{\rm{{0}}}$=5, and a ratio of internal over rest-mass energy
$\eta_0$=80.  Even though it would be important to check the
dependence of the results on all these parameters, as well as on the
progenitor mass and structure, the challenging nature of the
simulations prevented us from performing a thorough study of the
parameter dependencies.  The overall activity time of the central
engine was set taking into account that the width of the observed
duration of long BATSE gamma-ray bursts is mostly accounted for by an
engine lasting 20~s \citep{laz13b}. The variable temporal behavior of
the jet consisted of a series of N step functions (all with the same
``on'' and ``off'' durations, $\delta t$) before the engine luminosity
was suddenly decreased to a negligible value and eventually turned off
when the maximum integration time was reached. The two models
consisted of forty half-second episodes before the engine was turned
off. The active stages of the first model (m1) had a luminosity equal
to $L_0=5.33 \times$10$^{50}$~erg~s$^{-1}$, while the quiescent stages
were three orders of magnitude dimmer. The second model (m2) had a
monotonically decreasing step function distribution for the value of
the luminosity of the pulses. In order for both models to have the
same overall energy the initial pulse of model m2 had luminosity $1.95
L_0$, each subsequent pulse diminished 5\% until the final pulse had
luminosity $0.05 L_0$. In addition, we ran an extra simulation with a
single twenty second active period before the jet was turned off
(model m3, our control case). The characteristics of our models are
shown in Figure~\ref{fig:fig1}.  We must note that this study was
two-dimensional and one could expect that there would be significant
changes in three-dimensions. However, two-dimensional simulations are
much less demanding in terms of CPU hours and these simulations could
not be performed in 3D. A comparison between 2 and 3D simulations on
smaller domains showed that 2D simulations give reliable results since
the overall jet behavior is only marginally affected by the
dimensionality. \citet{lc13} show that the overall morphology and
large-scale features of 2D GRB jets resemble those from
three-dimensional, even though 2D jet models break out at later times
and present less turbulence.

\section{Results and Discussion}\label{sec:results}
\subsection{Global morphology. }\label{sec:morph}
\begin{figure*}
\centering
    \includegraphics[width=0.68\textwidth]{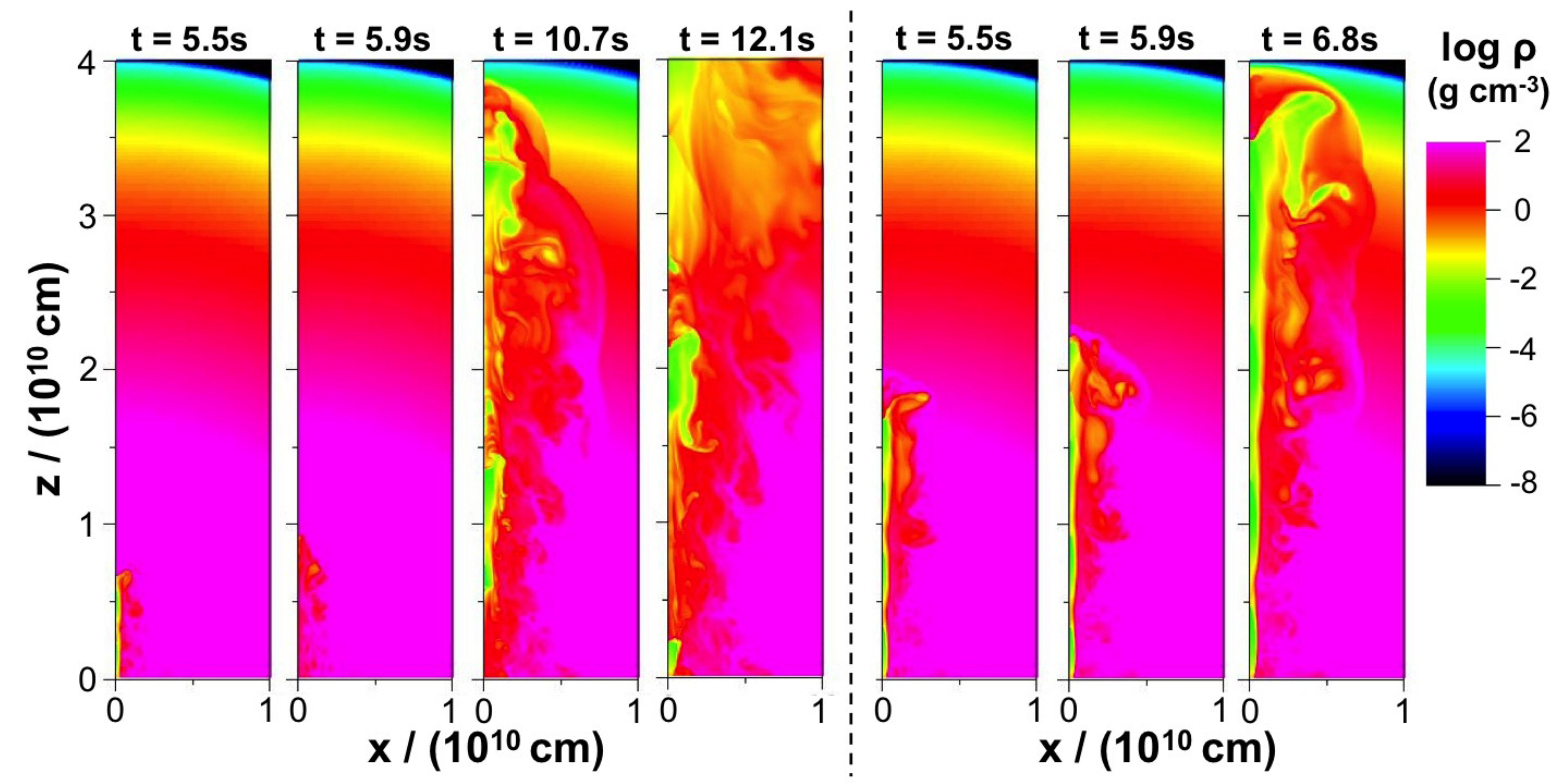}
    \caption{Density (g cm$^{-3}$) stratification maps for model m1 (4 left
      panels), and for model m3 (3 right panels) at different times.}
\label{fig:fig2}
\end{figure*}

Before we discuss the morphology, evolution, and photospheric emission
of the episodic jet, we first describe our control case (model m3),
the single 20 second spike model. The results for the single spiked
model were consistent with the results obtained from previous two-
\citep{zwm03, zwh04, mor07, mor10, miz13}, and three-dimensional
GRB-jet studies \citep{lc13}, as well as within the breakout time
range of \citet{brom11} analytical model. The 20 second single pulse
model drills through the stellar envelope and breaks out of the
progenitor (see the three right panels from
Figure~\ref{fig:fig2}). The forward, reverse, collimation and oblique
shocks are present, and the break out time was
$t_{\rm{{bo}}}$=6.8~s. We must point out that due to the fact that we
have a finer cocoon resolution, the breakout time of our control case
is slightly longer than that seen for example in \citep{mor10}. The
jet is at all times low-density ($\rho \sim$10$^{-3}$~g~cm$^{-3}$),
and before the jet breaks out of the star it is mildly
relativistic. The regions where the collimation and oblique shocks are
present reach $\Gamma$ values close to $\sim10$ within seconds (see
the schematic Figure 3 from \citet{miz13} for more details). Before
the jet breaks out of the star, these regions reach values close to
$\Gamma \sim$20. Once the jet breaks out of the star it reaches
$\Gamma \sim$130 (at $Z_{\rm{obs}}$).

Each pulse from the step jet models (m1 and m2) were low-density at
all times ($\rho \sim$10$^{-3}$~g~cm$^{-3}$) and mildly relativistic
before the break out ($\Gamma \sim$10). The quiescent periods between
the pulses, on the other hand, were at least two orders of magnitude
denser than the pulses and were sub-relativistic. The break out time
for model m1 and m2 was $t_{\rm{ {bo}}}$=10.7~s and 7.8~s,
respectively. Note that the break out time decreases monotonically
with increasing average engine luminosity, as expected. For both
episodic models, a set of the initial pulses is destroyed by the dense
stellar envelope before the jet is able to brake out of the
progenitor. For model m1 (m2) the first nine (seven) pulses are
destroyed. In order to clarify this the destruction of one of the
pulses is shown in the two left panels from
Figure~\ref{fig:fig2}. Notice how the fifth pulse from model m1
(present at 5.5~s) disappears before the next pulse is even
launched. Each time a pulse was engulfed by the stellar envelope, the
subsequent pulse managed to drill further out of the star and, unless
it reached the stellar surface, it would also be destroyed. Once the
jet brakes out of the star, the pulse destruction ceased to
occur. This is further clarified by the middle panels of
Figure~\ref{fig:fig2}. In these, when $t=10.7$~s the tenth pulse from
model m1 is not destroyed and reaches the stellar surface. The
subsequent pulses also break out of the star. For example, the
eleventh pulse (situated at $\sim$10$^{10}$~cm at 10.7s) when $t=12.1$
also breaks out of the progenitor. A similar behavior was observed in
simulations of a rapidly varying jets studied by \citet{mor10}. There,
jets with 0.1s pulses and random power variability on 0.1~s timescales
were used. In both cases, the first few seconds of jet variability
were wiped out by strong interaction with the stellar envelop as the
jet propagated. At later times, however, there was a strong
correspondence between the input variability of the central engine and
the short time scale variability of the jet seen at large radii. It is
noteworthy to mention that if the quiescent period between pulses was
too large, then the pressure from the stellar envelope would fill the
funnel that the recently destroyed pulse had created. For the chosen
progenitor, if the dormant epochs lasted more than one second then the
funnel would always collapse and the jet propagation would be
hindered.

\begin{figure*}
    \includegraphics[width=0.68\textwidth]{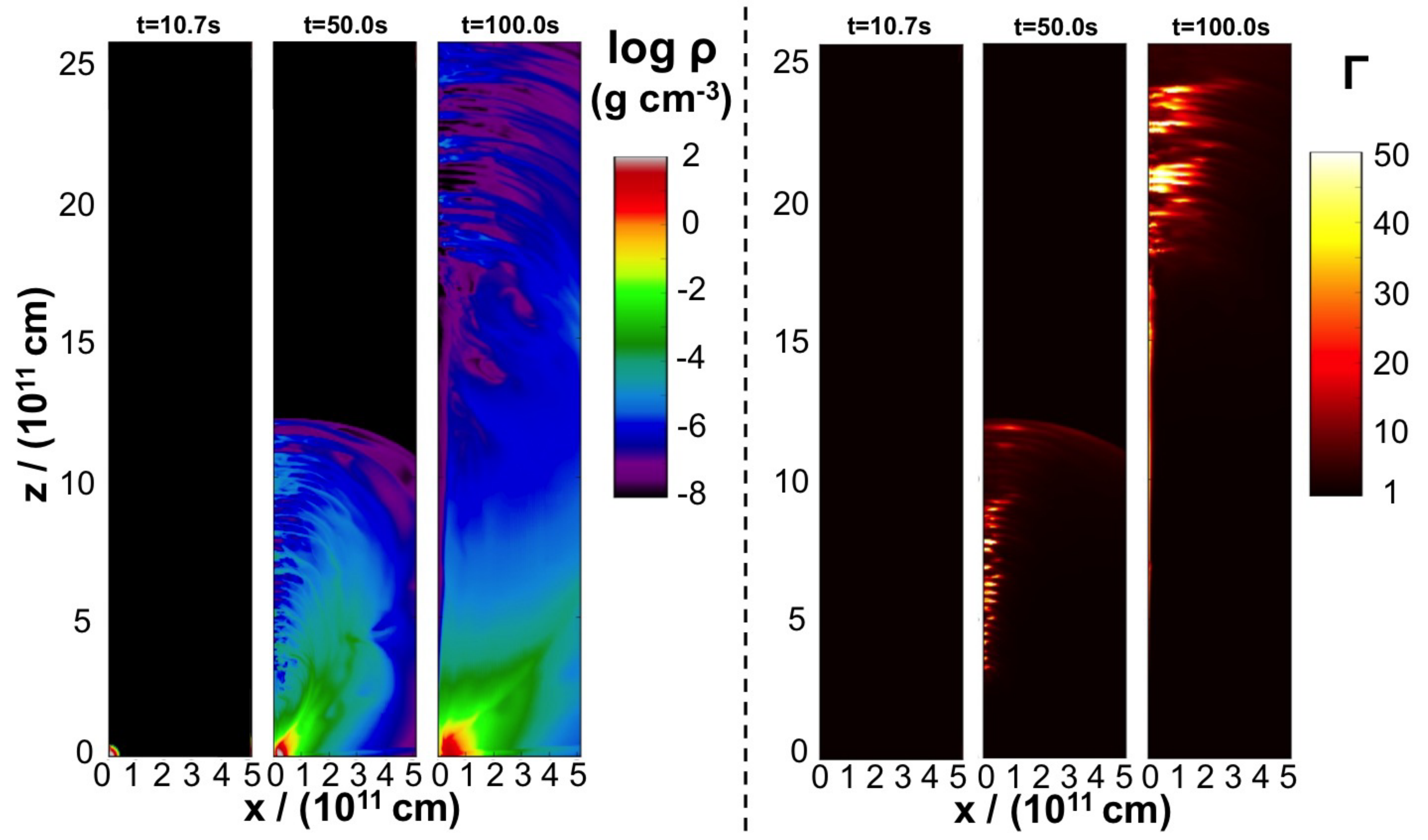}
    \caption{Density (g cm$^{-3}$) stratification maps (3 left
      panels), and Lorentz factor maps (3 right panels)
      for model m1.}
\label{fig:fig3}
\end{figure*}

The evolution of the variable jet through the ISM is illustrated in
Figure~\ref{fig:fig3}. In this, the density and $\Gamma$
stratification maps for model m1 at different times are shown. Both
models broke out of the star, and the episodic jets reach
$Z_{\rm{obs}}$ after approximately 100~s. Akin to the single spike,
before the break up time the active periods have $\rho
\sim$10$^{-3}$~g~cm$^{-3}$ and $\Gamma \sim$15. We must state that
model m2 had an overall evolution similar to that of m1. It mainly
differs in the fact that, since the initial pulses were more energetic
than those from m1, it broke out of the progenitor and reached
$Z_{\rm{obs}}$ faster than for m1. Also, as the control model, once
$t>t_{\rm{bo}}$ the active periods accelerate while decreasing their
density. The pulses reach values as low as $\rho
\sim$10$^{-6}$~g~cm$^{-3}$ (the quiescent epochs remain two orders of
magnitude lower), and Lorentz factors as high as $\Gamma$=80
(quiescent periods as expected have $\Gamma \sim$1).

\begin{figure}
\centering
    \includegraphics[width=\columnwidth]{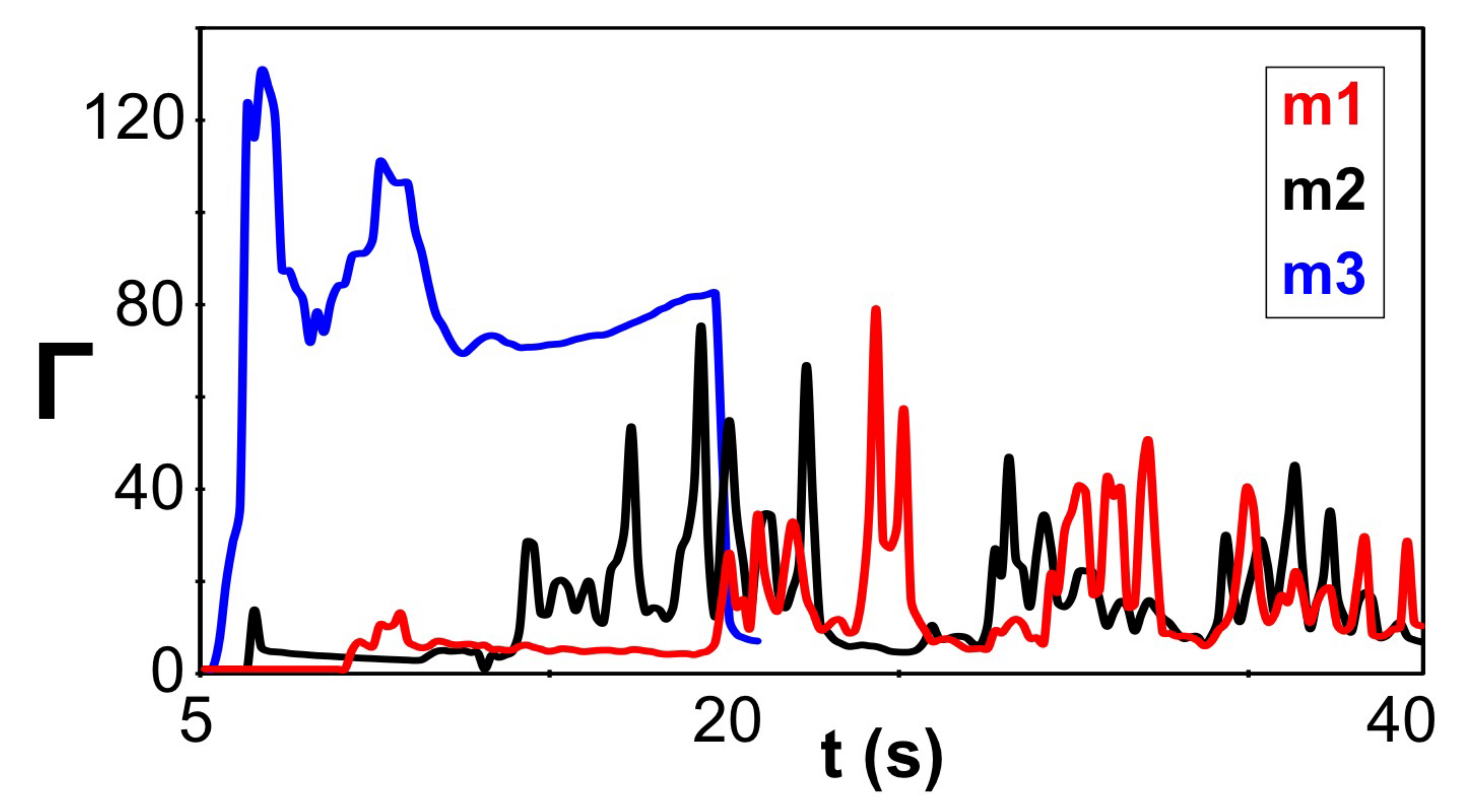}
    \caption{Temporal Lorentz factor evolution at $Z_{\rm{obs}}$ for models m1 (black), m2 (red), and m3 (blue).}
\label{fig:fig4}
\end{figure}

In order to illustrate the relativistic motion of the pulses,
Figure~\ref{fig:fig4} shows the temporal evolution of $\Gamma$ for an
observer set at $Z_{\rm{obs}}$ with a $\theta$=1$^o$ viewing
angle. The single pulse model initially reaches a Lorentz factor value
close to 120 and after $\sim$10~s remains fairly constant with
$\Gamma\sim$80. On the other hand, the episodic jet models behavior is
clearly present at $Z_{\rm{obs}}$ and oscillates between
$\Gamma=10-80$. The $\Gamma$ temporal structure for different viewing
angles for the episodic models also shows this variability, but the
maximum Lorentz values are noticeably lower (e.g. 56 and 34 for model
m1 seen at $Z_{\rm{obs}}$ with 3$^o$ and 5$^o$ viewing angle,
respectively).

\subsection{Photospheric light curves}\label{sec:phot}
\citet{giannios12} has shown that the spectrum in a relativistic outflow
is formed inside the photosphere, when the scattering optical depth
reaches a critical value 
\begin{equation}
\tau_T=46\frac{L_{53}^{1/6}f_\pm^{1/3}}{\Gamma_{2.5}^{1/3}\epsilon^{1/6}\eta_{2.5}^{1/3}}
\end{equation} 

The location of the critical optical depth is found by computing the
optical depth backward in space and time from an imaginary observed
located at $Z_{\rm{obs}}=2.5\times10^{12}$~cm from the center of the
explosion. Retarded time and the effect of the relativistic expansion of
the fluid are taken into account by enforcing \citep{laz13a}:
\begin{eqnarray}
&&46\frac{L_{53}^{1/6}f_\pm^{1/3}}{\Gamma_{2.5}^{1/3}\epsilon^{1/6}\eta_{2.5}^{1/3}}=
\nonumber \\ 
=-&&\!\!\!\!\!\!\!\!\!\int_{Z_{\rm{obs}}}^{Z_{\rm{sp}}(x)}
\sigma_T n^\prime\left(t_{\rm{obs}}-\frac{Z_{\rm{obs}}-z}{c},x,z\right)\,
\Gamma\left[1-\beta\cos(\theta_v)\right]\,dz \;\;\;\;
\end{eqnarray}
where $\beta\equiv\beta(t_{\rm{lab}},x,z)$ is the local velocity of
the outflow in units of the speed of light,
$\Gamma\equiv\Gamma(t_{\rm{lab}},x,z)$ is the local bulk Lorentz
factor, and $\theta_v\equiv\theta_v(t_{\rm{lab}},x,z)$ is the angle
between the velocity vector and the direction of the line of
sight. $x$ is the coordinate perpendicular to the line of sight, while
$z$ is the coordinate along the line of sight\footnote{Note that,
  strictly speaking, we do not find the spectral radius but the
  z-coordinate of the location at which the spectrum is formed. This
  allows us to implicitly take into account the equal arrival time
  surfaces.}. All the values of $\beta$, $\Gamma$, and $\theta_v$ are
evaluated at the same delayed coordinate
$(t_{\rm{lab}},x,z)\equiv\left(t_{\rm{obs}}-\frac{Z_{\rm{obs}}-z}{c},x,z\right)$
as the comoving density.

\begin{figure}
\centering
    \includegraphics[width=\columnwidth]{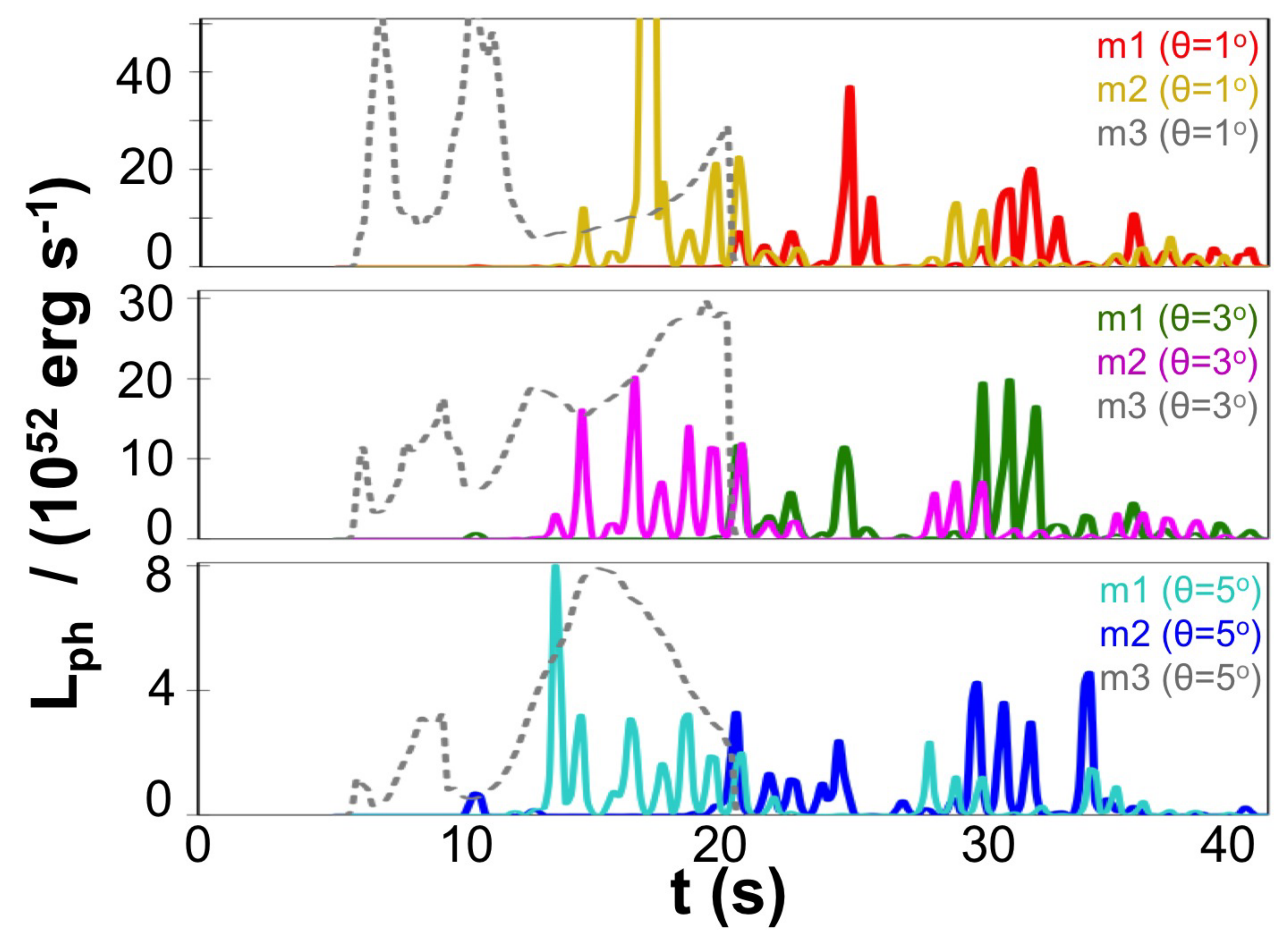}
    \caption{Photospheric light curves of the three models at
      different viewing angles (top panel is for $\theta$=1$^{o}$,
      middle panel $\theta$=3$^{o}$, and bottom panel
      $\theta$=5$^{o}$).}
\label{fig:fig5}
\end{figure}

Once the location of the spectral radius is found, we compute the
bolometric luminosity and the peak frequency of the emission following
\citet{laz13a}. We stress that we do not compute the whole spectral
shape of the burst, since that would require detailed radiation
transfer calculations beyond the scope of this research. The resulting
light curves are shown in Figure~\ref{fig:fig5}.

The variable behavior of models m1 and m2, as well as the observer
viewing angle, have direct impact on the photospheric light curves.
The photospheric light curve of the pulsed models also has an episodic
behavior. Though there is not a clear one to one relationship between
the launched pulses (shown in Figure~\ref{fig:fig1}) and the spikes in
the light curve (shown in Figure~\ref{fig:fig5}), it is clear that the
half second pulses produce $\approx$1~s episodes, a time lapse which
is actually the typical pulse width observed in the light curves of
variable long GRBs \citep{fenimore95, norris96}, and with the
luminosity ranging from 10$^{52}$erg~s$^{-1}$ to
4$\times$10$^{53}$erg~s$^{-1}$. This behavior is clearly not present
in the single 20~s pulse. The episodic temporal evolution is present
independently of the viewing angle, but as expected the luminosity
decreases for larger viewing angles. For example, for a 3$^{o}$
viewing angle the maximum value of the photospheric luminosity is
2$\times$10$^{53}$erg~s$^{-1}$, and 8$\times$10$^{52}$erg~s$^{-1}$ for
when the viewing angle is 5$^{o}$ (see middle and lower panels from
Figure~\ref{fig:fig5} respectively).  The different temporal
distributions with which the pulses from model m1 (pulses all with the
same luminosity) and m2 (pulses with a linearly decreasing step
function luminosity) were launched are not immediately recognizable in
the photospheric luminosity.  We computed the power spectra of all
light curves. The power spectra of the variable models (m1 and m2) are
indistinguishable from each other at all viewing angles, with a red
noise component that can be modeled as a $\nu^{-2}$ power law plus a
marked periodicity at ~1 Hz. This is not surprising since the engine
is strictly periodic with a 1 s period (fast variability is known to
propagate with the jet unaffected by the star: \citet{mor10}). The
non-variable model (m3) has a similar behavior but does not have any
periodic signal.  The power spectra of models m1 and m2 (the variable
ones) are not consisted with observed GRBs, since observed light
curves do not have any periodicity. However, given the lack of an easy
recipe for simulating a GRB light curve, any choice of the engine
luminosity profile would be as questionable as any other. We adopted a
periodic input in order to limit the degrees of freedom in the choice
of the engine variability.

\subsection{The Golenetskii correlation}\label{sec:gol}
\begin{figure}
\centering
    \includegraphics[width=\columnwidth]{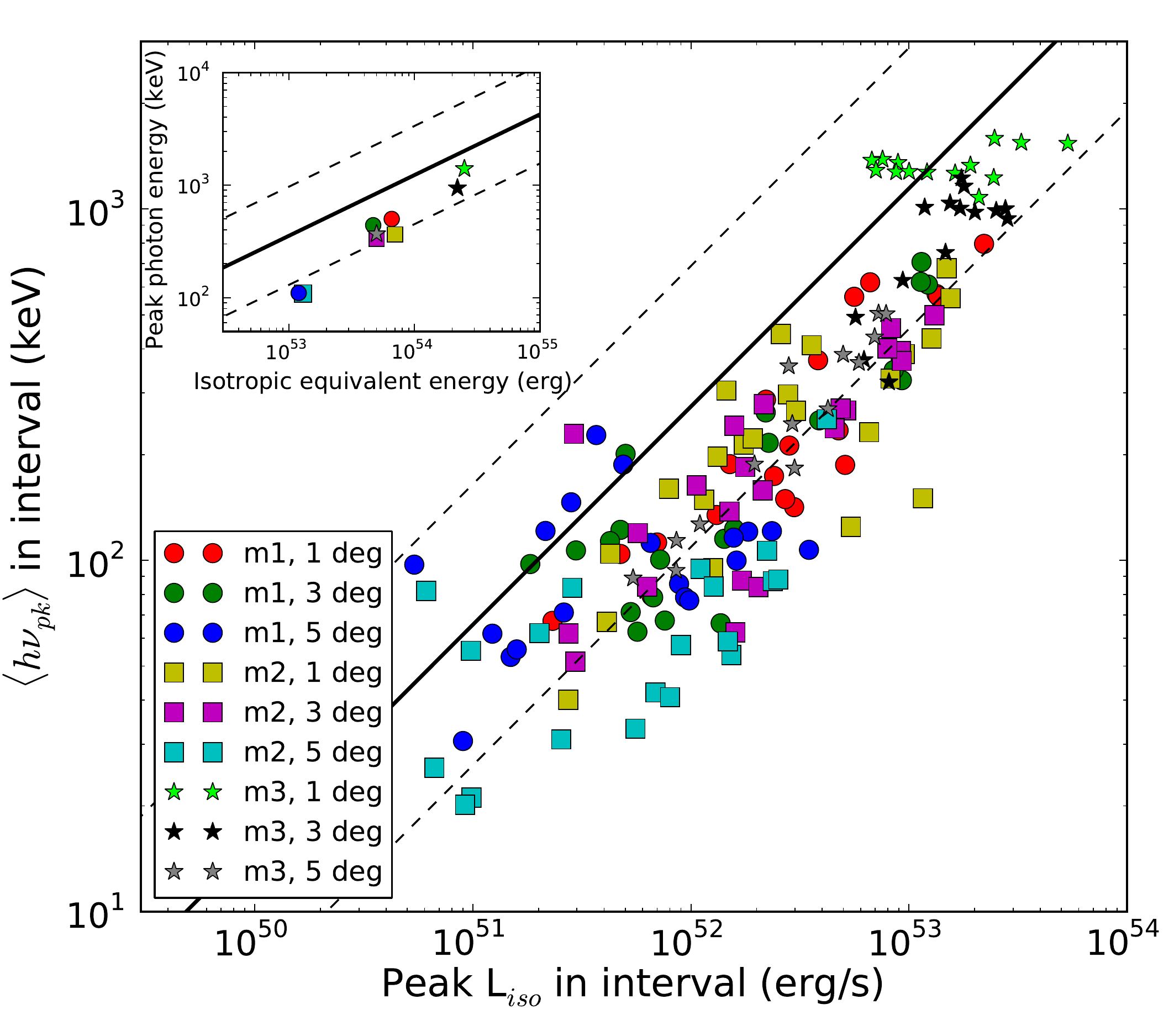}
    \caption{Synthetic Golenetskii correlation for models m1, m2 and m3 at
      various viewing angles, as indicated in the legend. The solid
      line represent the best fit of Fermi data \citep{lu12} and
      their 2-sigma confidence (dashed lines). The inset show the
      synthetic bursts on the Amati plane.}
\label{fig:fig6}
\end{figure}

The peak frequency (E$_{\rm{pk}}$) and corresponding isotropic
luminosity (L$_{\rm{{iso}}}$) were calculated for every half second
interval in which the engine was active. To better reproduce
observations, where the peak frequency is computed by time integrating
the spectrum over the interval, we report the luminosity-weighted
average peak frequency:
\begin{equation}
\langle h\nu_{\rm{pk}}\rangle_{\rm{int}}=h\frac{\int_t^{t+\delta t} L(t)\nu_{\rm{pk}}(t)\,dt}{\int_t^{t+\delta t} L(t)\,dt}
\end{equation}
The resulting relationship between the peak frequency and luminosity
for each burst, as well as the Amati relationship are shown in
Figure~\ref{fig:fig6}. The observational correlation between peak
frequency and luminosity was discovered by \citet{golenetskii83} and
is also known as the internal Yonetoku correlation \citep[i.e. the
Yonetoku correlation for individual bursts,][]{yonetoku10}. We show
the observed best fit correlation with a solid line \citep{lu12} and
we delimit the 2-sigma confidence region with dashed lines.  For all
three models the data from the synthetic light curves and spectra show
agreement with the observations. As can be seen in
Figure~\ref{fig:fig6} the majority of the dataset we obtain is within
the two-sigma confidence region. Specifically, the Golenetskii
relationship for the whole dataset
(log(E$_{\rm{pk}}$=-28.60$\pm$1.40)+(0.59$\pm$0.03)) is very similar
to that from the bright variable GRB dataset observed by Fermi
\citep{lu12}
(log(E$_{\rm{pk}}$=-29.854$\pm$0.178)+(0.621$\pm$0.003)). However,
when the models are considered individually, the agreement
weakens. For example, the average correlations for all viewing angles
of the three models presented are:
log(E$_{\rm{pk}}$=-20.60$\pm$2.10)+(0.44$\pm$0.04),
log(E$_{\rm{pk}}$=-24.10$\pm$2.50)+(0.50$\pm$0.05), and
log(E$_{\rm{pk}}$=-35.30$\pm$3.00)+(0.70$\pm$0.06) for m1, m2, and m3,
respectively.

The synthetic data populate the region in the lower right of the
observed correlation and are therefore typical of a burst that is
somewhat ``soft'' for the given luminosity. In the inset of
Figure~\ref{fig:fig6} we show where the time-integrated pulsed bursts
would lie in the Amati plane.  Such bursts populate the lower right
part of the Amati correlation and are also within the two-sigma
confidence region. The Amati relationship obtained (E$_{\rm{pk}}$=(3.3
E$_{iso,52}^{0.79}$)) is quite steeper than the \citet{amati02}
relationship (E$_{\rm{pk}}$=(118 E$_{iso,52}^{0.486}$)).  The fact
that the simulated bursts populate the lower right portion of the
Amati and Golenetskii correlations is worth further investigation. A
previous set of simulations of constant engine GRBs showed
quantitative agreement with the Amati correlation \citep{laz13a}. At
this stage, it is unclear if the difference is due to the different
numerical setup or to the somewhat extreme variability adopted in this
paper, with the engine shutting off for half a second periods
regularly. We are planning to perform a more thorough study of
variable jets with different variability properties and inside
different progenitors to ascertain if the cause of the softness is
intrinsic, numerical, or related to the particular parametrization of
the jet/star pair used in this paper.  One important note about
Figure~\ref{fig:fig6} is that the graph shows only the properties of
the bright pulses, i.e., those with a peak luminosity larger than 1
per cent of the brightest peak. This choice was motivated by the fact
that observationally only the brightest part of a burst is amenable to
time-resolved spectral analysis and can be shown in the Golenetskii
plane. Since simulated light curves have no noise, we can measure
luminosity and peak frequencies for the weakest burst
intervals. Interestingly, these intervals deviate from the correlation
shown in the figure adding a flatter tail, lying above the
correlation, in the range of luminosities
$10^{48}<L_{\rm{iso}}<10^{50}$~erg/s. It is unclear at this moment
whether this is a model prediction or an artifact of the
simulations. As a matter of fact, time intervals with low-luminosity
are characterized by an enhanced baryon loading in the outflow and our
spectral calculations are not fully trustable when the location of the
spectral radius approaches the edge of the simulation domain.

\section{Summary and Conclusions}\label{sec:conc}
We presented the results of a set of numerical simulations of
long-duration GRB jets followed as they propagate through their
progenitor star, reach break-out, and expand outward until reaching
the radius at which the spectrum of the advected radiation converge to
a constant shape to be eventually released at the
photosphere. This is the first time that jets from engines with
variable luminosity are studied in such an extended domain. Our
simulations allow us to explore whether the photospheric emission of
jets from unsteady engines follows the Golenetskii correlation between
the time-resolved luminosity and spectral peak. 

We find that the synthetic light curves and spectra from our three
models reproduce the Golenetskii and Amati correlations, lending more
support to the scenario in which the bulk of the burst prompt
radiation is advected in the outflow and released at the
photosphere. One notable exception is the light curve for an on-axis
observer in our control model (m3). In that case, the synthetic
datapoint for an horizontal line that does not follow the Golenetskii
correlation. It appears, therefore, that while all GRBs from variable
engines obey the correlation, outliers can be produced by engines of
constant luminosity. Our simulations attempted to reproduce the
correlation as a pulse-by-pulse phenomenon. The engine was set up to
produce short pulses of half a second. Half a second features were
detected in the light curves and analyzed as pulses, each pulse
providing a single point in the Golenetskii plane. Observationally,
the latter correlation is detected also within pulses, i.e., when the
signal to noise is so high that a pulse can be split in
sub-intervals. Our simulations cannot address this situation, at
present. A new set of simulations with longer pulses and higher
temporal resolution is planned and will be presented in a future
publication.

 \textbf{Acknowledgements}
 We would like to thank the anonymous referee for constructive
 suggestions that led to the improvement of this paper.  We thank
 S.E. Woosley and A. Heger for making their pre-SN models
 available. The software used in this work was in part developed by
 the DOE-supported ASC/Alliance Center for Astrophysical Thermonuclear
 Flashes at the University of Chicago. This work was supported in part
 by the Fermi GI program grant NNX12AO74G and Swift GI program grant
 NNX13A095G (DL and DLC). BJM is supported by an NSF Astronomy and
 Astrophysics Postdoctoral Fellowship under award AST-1102796.


\label{lastpage}
\end{document}